%% file: main.tex
\documentclass[a4paper,11pt]{article}
\usepackage{jinstpub}

\title{Lowering threshold of NaI(Tl) scintillator to 0.7 keV in the COSINE-100 experiment}

\collaboration[c]{on behalf of COSINE-100 collaboration}

\input{authors}

\abstract{
COSINE-100 is a direct dark matter search experiment, with the primary goal of testing the annual modulation signal observed by DAMA/LIBRA, using the same target material, NaI(Tl). In previous analyses, we achieved the same 1\,keV energy threshold used in the DAMA/LIBRA’s analysis that reported an annual modulation signal with 11.6$\sigma$ significance.
In this article, we report an improved analysis that lowered the threshold to 0.7\,keV, thanks to the application of Multi-Layer Perception network and a new likelihood parameter with waveforms in the frequency domain. The lower threshold would enable a better comparison of COSINE-100 with new DAMA results with a 0.75\,keV threshold and account for differences in quenching factors. Furthermore the lower threshold can enhance COSINE-100's sensitivity to sub-GeV dark matter searches. 
}


\arxivnumber{2408.14688} 

\begin{document}
\maketitle
\flushbottom

\input{intro}
\input{cosine}
\input{sec_para}

\input{sec_pure}

\input{sec_learning}
\input{conclusion}

\acknowledgments
We thank the Korea Hydro and Nuclear Power (KHNP) Company for providing underground laboratory space at Yangyang and the IBS Research Solution Center (RSC) for providing high performance computing resources. 
This work is supported by:  the Institute for Basic Science (IBS) under project code IBS-R016-A1,  NRF-2021R1A2C3010989, NRF-2021R1A2C1013761 and RS-2024-00356960, Republic of Korea;
NSF Grants No. PHY-1913742, United States; 
STFC Grant ST/N000277/1 and ST/K001337/1, United Kingdom;
Grant No. 2021/06743-1, 2022/12002-7 and 2022/13293-5 FAPESP, CAPES Finance Code 001, CNPq 304658/2023-5, Brazil;
UM grant No. 4.4.594/UN32.14.1/LT/2024, Indonesia.

\bibliographystyle{JHEP}
\bibliography{biblio.bib}

\end{document}

%% file: authors.tex
\author[a,b]{G.~H.~Yu}
\author[c]{N.~Carlin}
\author[b,d]{J.~Y.~Cho}
\author[e,b]{J.~J.~Choi}
\author[e]{S.~Choi}
\author[f]{A.~C.~Ezeribe}
\author[c]{L.~E.~Fran\c{c}a}
\author[g]{C.~Ha}
\author[h,i,j]{I.~S.~Hahn}
\author[k]{S.~J.~Hollick}
\author[b,j]{E.~J.~Jeon}
\author[e]{H.~W.~Joo}
\author[b]{W.~G.~Kang}
\author[l]{M.~Kauer}
\author[b]{B.~H.~Kim}
\author[d]{H.~J.~Kim}
\author[g]{J.~Kim}
\author[b]{K.~W.~Kim}
\author[b]{S.~H.~Kim}
\author[e]{S.~K.~Kim}
\author[j,b]{W.~K.~Kim}
\author[b,j]{Y.~D.~Kim}
\author[b,j]{Y.~H.~Kim}
\author[b,1]{Y.~J.~Ko\note{Corresponding author.}}
\author[d]{D.~H.~Lee}
\author[b]{E.~K.~Lee}
\author[j,b]{H.~Lee}
\author[b,j]{H.~S.~Lee}
\author[h]{H.~Y.~Lee}
\author[b]{I.~S.~Lee}
\author[b]{J.~Lee}
\author[d]{J.~Y.~Lee}
\author[b,j]{M.~H.~Lee}
\author[b,j]{S.~H.~Lee}
\author[e]{S.~M.~Lee}
\author[g]{Y.~J.~Lee}
\author[b]{D.~S.~Leonard}
\author[d]{N.~T.~Luan}
\author[c]{V.~H.~A.~Machado}
\author[c]{B.~B.~Manzato}
\author[k]{R.~H.~Maruyama}
\author[f]{R.~J.~Neal}
\author[b]{S.~L.~Olsen}
\author[j,b]{B.~J.~Park}
\author[m]{H.~K.~Park}
\author[n]{H.~S.~Park}
\author[o]{J.~C.~Park}
\author[b]{K.~S.~Park}
\author[d]{S.~D.~Park}
\author[c]{R.~L.~C.~Pitta}
\author[p]{H.~Prihtiadi}
\author[b]{S.~J.~Ra}
\author[a,q]{C.~Rott}

\author[b]{K.~A.~Shin}
\author[c]{D.~F.~F.~S. Cavalcante}
\author[o]{M.~K.~Son}
\author[f]{N.~J.~C.~Spooner}
\author[d]{L.~T.~Truc}
\author[r]{L.~Yang}

\affiliation[a]{Department of Physics, Sungkyunkwan University, Suwon 16419, Republic of Korea}
\affiliation[b]{Center for Underground Physics, Institute for Basic Science (IBS), Daejeon 34126, Republic of Korea}
\affiliation[c]{Physics Institute, University of S\~{a}o Paulo, 05508-090, S\~{a}o Paulo, Brazil}

\affiliation[d]{Department of Physics, Kyungpook National University, Daegu 41566, Republic of Korea}
\affiliation[e]{Department of Physics and Astronomy, Seoul National University, Seoul 08826, Republic of Korea}
\affiliation[f]{Department of Physics and Astronomy, University of Sheffield, Sheffield S3 7RH, United Kingdom}
\affiliation[g]{Department of Physics, Chung-Ang University, Seoul 06973, Republic of Korea}
\affiliation[h]{Center for Exotic Nuclear Studies, Institute for Basic Science (IBS), Daejeon 34126, Republic of Korea}
\affiliation[i]{Department of Science Education, Ewha Womans University, Seoul 03760, Republic of Korea}
\affiliation[j]{IBS School, University of Science and Technology (UST), Daejeon 34113, Republic of Korea}
\affiliation[k]{Department of Physics and Wright Laboratory, Yale University, New Haven, CT 06520, USA}
\affiliation[l]{Department of Physics and Wisconsin IceCube Particle Astrophysics Center, University of Wisconsin-Madison, Madison, WI 53706, USA}
\affiliation[m]{Department of Accelerator Science, Korea University, Sejong 30019, Republic of Korea}
\affiliation[n]{Korea Research Institute of Standards and Science, Daejeon 34113, Republic of Korea}
\affiliation[o]{Department of Physics and IQS, Chungnam National University, Daejeon 34134, Republic of Korea}
\affiliation[p]{Department of Physics, Universitas Negeri Malang, Malang 65145, Indonesia}
\affiliation[q]{Department of Physics and Astronomy, University of Utah, Salt Lake City, UT 84112, USA}
\affiliation[r]{Department of Physics, University of California San Diego, La Jolla, CA 92093, USA}

\emailAdd{tksxk752@naver.com}
\emailAdd{yjko@ibs.re.kr}

%% file: intro.tex
\section{Introduction}
\label{sec:intro}
The existence of dark matter~(DM) has been supported by decades of astrophysical and cosmological observations, with DM estimated to make up about 27\% of the universe~\cite{Darkmatter:2006eq, Planck:2013pxb}. Many theoretical models were developed to describe DM particles and their interactions with Standard Model particles. One of the leading candidates is a weakly interacting massive particle~(WIMP), which can satisfy the conditions outlined by cold DM and yields the proper relic density for DM abundance~\cite{Lee:1977ua,Jungman:1995df}. Despite efforts of many underground experiments to search for DM direct interactions, no dark matter signals have been discovered except for a putative claim by DAMA/LIBRA.~\cite{Bernabei:2013xsa,Bernabei:2018jrt,PhysRevD.98.030001}

DAMA/LIBRA has claimed a DM annual modulation signature due to the changing velocity of the Earth in the galactic reference frame. The regions allowed by the DAMA/LIBRA data with interpretations based on the standard halo model and various WIMP-nucleus interaction models have already been ruled out by several experiments~\cite{collaboration2018dark,amole2019dark}, notably by COSINE-100 using the same target material, NaI(Tl)~\cite{PhysRevD.98.030001,COSINE-100:2021xqn}. However, it remains important to conduct a model-independent annual modulation analysis, and there are global efforts to replicate the modulation signature from several NaI experiments~\cite{DM-Ice:2016snk,amare2021annual,adhikari2022three}.

The first results from DAMA/LIBRA-phase2 used an analysis energy threshold of 1\,keV down from 2\,keV in phase 1~\cite{Bernabei:2018jrt}. Recently, they published a modulation amplitude spectrum with a lower threshold of 0.75\,keV~\cite{Bernabei:2022xgg}, testing their results via a direct comparison requires the same energy calibration.
It is important to note that DAMA reconstructed the energy based on a linear calibration that does not account for nonproportionality~(nPR)~\cite{DAMA_calib}, whereas the energy used in this study does account for nPR. The energies corresponding to 1\,keV and 0.75\,keV in DAMA are equivalent to 1.25\,keV and 0.93\,keV, respectively, when nPR is taken into account~\cite{COSINEnPR}.

Since the nuclear recoil signal by WIMPs is what we are looking for in the search, the calibration for the nuclear recoil energy is a critical issue. This can be taken into account by using the measurements of the quenching factor~(QF), which is the ratio of the nuclear recoil energy to the electron equivalent energy. Recent QF measurements from COSINE-100~\cite{NaIQF,lee2024measurements} and other research groups~\cite{collar2013quenching,xu2015scintillation,stiegler2017study,PhysRevC.110.014613} show a different behavior than DAMA's measurements~\cite{Bernabei:1996vj}, notably that the QF at DAMA's threshold is only about half that of DAMA. Although the recent QF measurements from different crystals by different groups are consistent with each other, we cannot rule out the possibility that the DAMA crystals have different characteristics. When comparing nuclear recoil energy spectra, we conservatively apply the measured QFs by each experiment to their respective electron recoil energy spectra. Thus, to reach the same nuclear recoil energy threshold as DAMA will require a 50\% lower electron equivalent energy threshold for COSINE-100.

While the same threshold is not necessarily required for direct comparison, getting closer to the threshold of DAMA has the advantage of reducing the required exposure. This paper describes the event selection that enabled the 0.7\,keV energy threshold to be achieved.
Given that the nuclear recoil spectra of low-mass WIMPs decrease very steeply, their detection is highly sensitive to the energy threshold. The 0.7\,keV energy threshold is also expected to enhance the sensitivity of WIMP searches, especially in the low-mass region~\cite{COSINE-100:2020wrv}, or related to the Migdal effect~\cite{migdal1941ionization,Mashiro:JHEP2018,PhysRevD.105.042006}. 
Additionally, since both Na and I have odd proton numbers, the sensitivity to spin-dependent WIMP-proton interactions is expected to be competitive globally~\cite{Bednyakov:2006,PhysRevD.89.123521}.

Section~\ref{sec:experiment} provides an overview of the COSINE-100 experiment, while in Sec.~\ref{sec:parameter} we describe several parameters developed to separate scintillation events from noise events. In Sec.~\ref{sec:sample}, we cover the selection of pure scintillation event samples and Sec.~\ref{sec:event_selection} discuss the multivariate machine learning~(ML) technique used and the validation of the event selection. Finally, Sec.~\ref{sec:summary} summarizes the results and highlights implications for future research.

%% file: cosine.tex
\section{Experimental Setup of COSINE-100}
\label{sec:experiment}
COSINE-100~\cite{Adhikari:2017esn} is a direct DM search experiment installed at Yangyang Underground Laboratory~(Y2L) in South Korea~\cite{LEE2006201,Kim:2012rza} with 700\,meters of rock overburden~\cite{COSINE-100:2017jinst}. The detector consists of eight thallium-doped NaI crystals with a total mass of 106\,kg. Two 3-inch Hamamatsu R12669SEL photomultiplier tubes~(PMTs) are attached to both sides of each crystal, encapsulated in a copper case. 
``The effective light yield'' of the crystals, the ratio of the total number of photoelectrons generated in both PMTs to the energy deposited in the crystal, was estimated from the 59.6\,keV peak of $^{214}$Am and averaged 15\,photoelectrons per keV~\cite{Adhikari:2017esn}.

The crystals, arranged in a 4$\times$2 array, are immersed in a 2200\,L linear alkylbenzene-based liquid scintillator~(LS) which serves as an active veto for coincidence events between crystals and LS.~\cite{Jengsu:2017,Adhikari:2020asl}. The LS is contained in a 1\,cm thick acrylic box surrounded by 3\,cm thick copper. To reduce the background by radiation from external sources, a 20\,cm thick lead shield is placed on the next outer layer, and the outermost layer is equipped with plastic scintillator panels to identify cosmic muon events~\cite{COSINE-100:2017jinst,COSINE-100:MuonMod}.

Event tagging by the muon counter is the first step in event selection: once a muon event is identified, all events from crystals within a 30 ms veto window are rejected. Next, if a hit occurs in a crystal with a time-coincident hit in any other crystal or LS, the event is classified as a multiple-hit event and is excluded from the WIMP-search data. This process plays an important role in background reduction in the low energy region, especially by tagging 0.87 and 3.2\,keV X-ray events from $^{22}$Na and $^{40}$K, respectively~\cite{Adhikari:2020asl}.
Additionally, multiple-hit events can serve as scintillation samples because they are less likely to contain noise events due to the time coincidence condition with other events.

Each PMT signal is amplified in a preamplifier and then digitized by an analog-to-digital converter~(ADC) at a sampling rate of 500\,MHz. The ADC is connected to the trigger control board~(TCB)~\cite{COSINE-100:2018rxe}. When the signal is triggered by the TCB, it is recorded as a waveform within an 8\,$\mu$s time window, starting from 2.37\,$\mu$s before the trigger time. Based on waveform characteristics, such as mean time and pulse slope, pulse shape discrimination~(PSD) is utilized for event selection.

PSD can be used to improve the sensitivity of WIMP searches by separating noise events from scintillation events in the low-energy region. In low-energy data analysis below 2\,keV, scintillation events are overwhelmed by noise events. These noise events are generated by PMTs due to dark currents, radioactive contamination in the dynode circuitry, or occasional flashes~\cite{photonics2000photomultiplier}. Since the low-energy region overlaps with the region of interest for WIMP searches, it is critical to exclude these noise events from WIMP-search data.

Two widely used PSD parameters, the charge-weighted mean-time and the charge ratio for different integration ranges, are also being used in COSINE-100. In addition, a likelihood parameter was developed to utilize pulse shape information fully and already reached a threshold of 1\,keV through multivariate boosted decision tree~(BDT) training with discrimination parameters~\cite{COSINE-100:2020wrv}. To achieve the lower threshold of 0.7\,keV, several additional likelihood parameters have been developed. We will describe the event selection study in the following sections.

%% file: sec_para.tex
\section{Likelihood-Based PSD Parameters}
\label{sec:parameter}

\begin{figure}[tb]
    \centering
    \centerline{\includegraphics[width=0.95\linewidth]{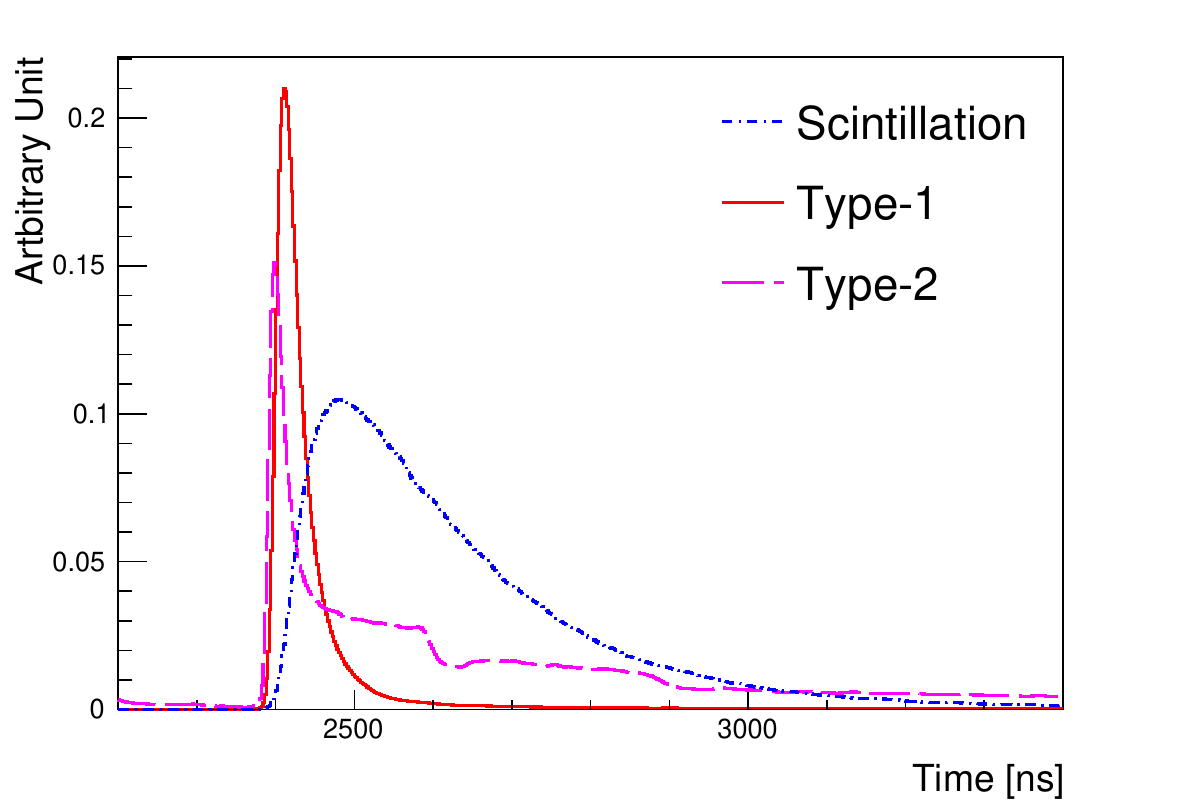}}
    \caption{Waveforms accumulated by event types in the 2 to 6\,keV energy region. The selection of event types followed the method in the previous 1\,keV threshold study~\cite{COSINE-100:2020wrv}.
    }
    \label{fig:waveform}
\end{figure}

The scintillation events in COSINE-100 are contaminated by PMT-induced noise. In Ref.~\cite{COSINE-100:2020wrv}, these noise events were classified as Type-1 and Type-2, based on the parameter space defined by charge-weighted mean-time and pulse-shape parameters. Type-1 noise refers to events with large spikes and short mean times, while Type-2 noise consists of events with small spikes and long mean times. This distinction is illustrated in Fig.~\ref{fig:waveform}. During the event selection process, Type-2 noise is more difficult to distinguish from scintillation events due to its long-tailed structure, which resembles that of the scintillation events.

In our previous study~\cite{COSINE-100:2020wrv}, we introduced a ``likelihood parameter ($p_l$)", which calculates and parametrizes the likelihood value between an event and the signal and noise templates shown in Fig.~\ref{fig:waveform}. This parameter was effective in separating Type-2 noise events and played a crucial role in achieving a 1\,keV threshold. However, this alone was insufficient to lower the threshold below the 1\,keV region. For more effective discrimination, we use individual PMT waveforms rather than crystal-based summed waveforms and different parameters for each noise type. We also apply new likelihood parameters using Fourier-transformed waveforms to further improve discrimination.

\subsection{PMT-based likelihood parameter}
\label{sec:PMT_parameter}
As an initial step to calculate the $p_l$, the scintillation signal events and the PMT noise events are collected separately to create a waveform template for each. The template is obtained by accumulating the waveforms from both PMTs attached to the crystal. The likelihood values of each event's crystal waveform with the templates are then calculated and combined to derive the $p_l$~\cite{COSINE-100:2020wrv}.

In this analysis, two aspects of the parameter have been improved. First, we use waveforms from each PMT individually instead of the summed waveform of both PMTs, computing the likelihood value per PMT because noise events are more likely to exhibit a large asymmetry between the two PMTs. Second, we create templates for different noise types and compute likelihood values separately for each noise type. These templates are depicted in Fig.~\ref{fig:waveform}.

The likelihood value for $i^\mathrm{th}$ PMT is defined as
\begin{equation}
\label{eq:likelihood_sig}
    \ln{\mathcal{L}_i} = \sum_j \left[ T_{ij} - W_{ij} + W_{ij} \cdot \ln{\frac{W_{ij}}{T_{ij}}} \right],
\end{equation}
where $T_{ij}$ and $W_{ij}$ denote the amplitudes at the $j^\mathrm{th}$ time bin of the template and event waveforms, after normalizing to the same integral size.
Depending on the template used, we can compute the likelihood values $\mathcal{L}^S_i$, $\mathcal{L}^{N1}_i$, and $\mathcal{L}^{N2}_i$ for scintillation signal, and Type-1 and Type-2 noise, respectively. Using these likelihood values, a PSD parameter for discriminating Type-1 noise events is defined as
\begin{equation}
    P_\mathcal{L}^{N1} = \frac{\sqrt{\ln{\mathcal{L}_1^{N1}}\cdot \ln{\mathcal{L}_2^{N1}}} - \sqrt{\ln{\mathcal{L}_1^{S}}\cdot \ln{\mathcal{L}_2^{S}}}}{2}.
\label{eq:plpar1}
\end{equation}
And the parameter $P_\mathcal{L}^{N2}$ for Type-2 noise is defined in a similar manner. The definition is somewhat arbitrary but is chosen in a way that effectively discriminates against noise. 

\begin{figure*}[tb]
    \centering
    \includegraphics[width=1.0\linewidth]{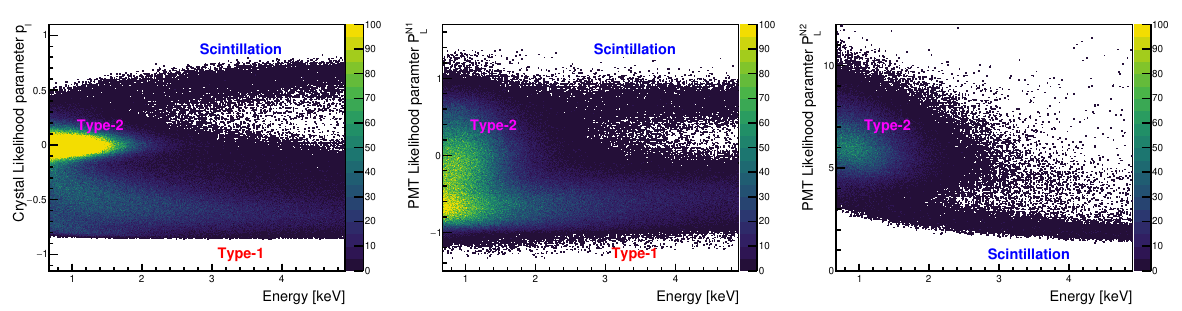}
    \caption{Two-dimensional distributions of each likelihood parameter versus energy with WIMP-search data. The left plot shows the crystal-based likelihood parameter $p_l$ developed in the previous study, while the middle and right plots show the PMT-based likelihood parameters $P_\mathcal{L}^{N1}$ and $P_\mathcal{L}^{N2}$, respectively. Type-1 noise events are removed in the right plot to illustrate a clear separation of Type-2 noise from scintillation events.}
    \label{fig:plpar}  
\end{figure*}

Figure~\ref{fig:plpar} shows two-dimensional distributions of the likelihood parameters versus energy. The middle plot is the distribution for $P_\mathcal{L}^{N1}$ versus energy, with Type-1 noise events distributed horizontally at the bottom, while scintillation events are distributed horizontally at the top, showing good discrimination power as intended. Meanwhile, one can see that the Type-2 noise events are mostly distributed at low energy and overlap with the region of the scintillation signal. Instead, they are well separated in the right plot, which shows the distribution for $P_\mathcal{L}^{N2}$ versus energy. Notably, it demonstrates better discrimination power at energies below 1\,keV when compared to $p_l$ in the left plot.

\subsection{FFT-based likelihood parameter}
\label{sec:FFT_parameter}
Given that the pulse shapes of scintillation and noise events exhibit distinct characteristics, interpreting these differences using Fourier transformation can be a valuable tool. Therefore, we apply the Fast Fourier Transformation~(FFT) to the waveforms and utilize the power spectra for the analysis.

\begin{figure}[htb]
    \centering
    \centerline{\includegraphics[width=0.9\linewidth]{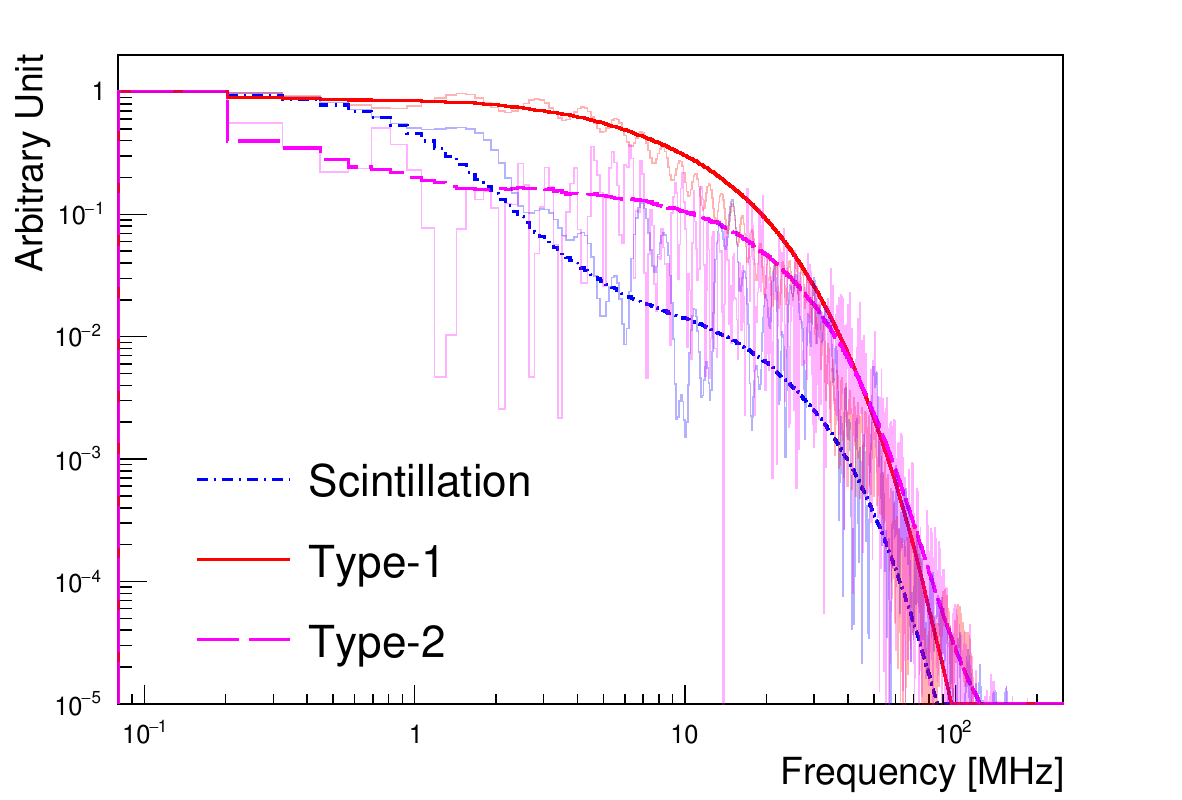}}
    \caption{Templates of FFT waveform for each event type. Each template is scaled to have value 1 in the first bin, which indicates the average value of the raw waveform. The selection criteria  followed the method in the previous 1\,keV threshold study~\cite{COSINE-100:2020wrv}. The histogram of a representative event for each event type is overlaid using the same color.}
    \label{fig:FFT}  
\end{figure}

To parametrize the features of power spectra, we make templates as shown in Fig.~\ref{fig:FFT}, and compute likelihood values for each event using Eq.~\ref{eq:likelihood_sig} as in Sec.~\ref{sec:PMT_parameter}. Here, the calculation is not based on each PMT but the summed waveform from two PMTs on the crystal.
The calculation range is set differently for each noise type : [0-6.14]\,MHz for Type-1 noise and,  [0-1.24]\,MHz for Type-2 noise. 
This is because the high-frequency region of the power spectrum contains less significant characteristics of waveform structure and has large event-by-event fluctuation.


The parameters for separating PMT-noise events with the FFT likelihood values have been defined as
\begin{equation}
\label{eq:flpar}
    F_\mathcal{L}^{N1}
    = \frac{\ln{\mathcal{L}_\mathrm{FFT}^{N1}} - \ln{\mathcal{L}_\mathrm{FFT}^{S}}}
    {\ln{\mathcal{L}_\mathrm{FFT}^{N1}} + \ln{\mathcal{L}_\mathrm{FFT}^{S}}}
\end{equation}
where $\mathcal{L}_\mathrm{FFT}^{N1}$ (or $\mathcal{L}_\mathrm{FFT}^{N2}$) and $\mathcal{L}_\mathrm{FFT}^{S}$ are the likelihood values using the Type-1 (or Type-2) PMT-noise and scintillation signal templates, respectively. It is important to note here that the parameters for Type-1 and Type-2 use likelihood values calculated with their respective ranges and templates.

\begin{figure*}[tb]
    \centering
    \centerline{\includegraphics[width=1.0\linewidth]{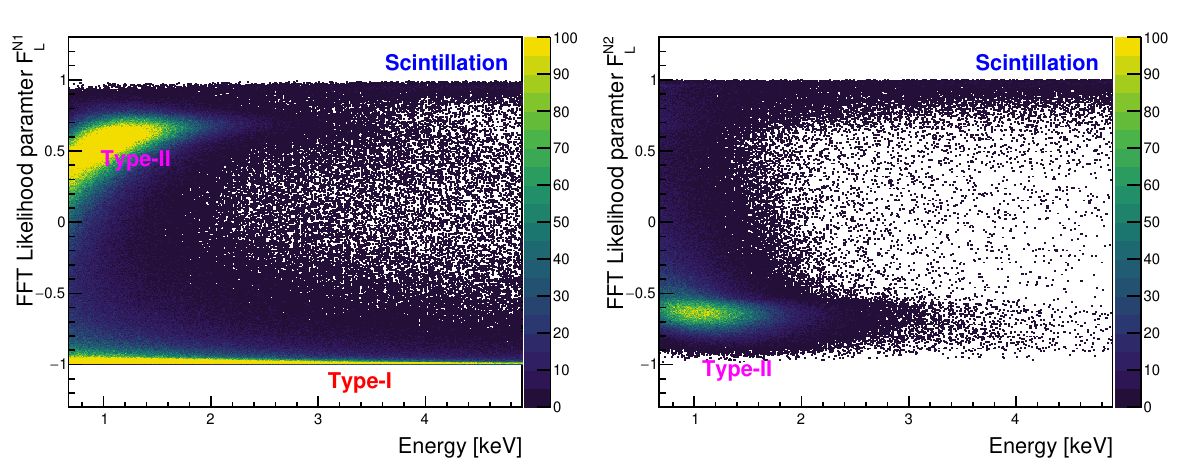}}
    \caption{
    Two-dimensional distributions of FFT likelihood parameter versus energy with WIMP-search data. The left plot shows the FFT parameter for Type-1 noise while the right plot shows that for Type-2 noise. Type-1 noise events are removed in the right plot for clear illustration purposes.
    }
    \label{fig:flpar}  
\end{figure*}

Figure~\ref{fig:flpar} shows the two-dimensional distributions between the FFT likelihood parameters and energy. As intended, the parameters demonstrate strong discrimination power for the each type of noise. They are input to multivariate ML with other PSD parameters, including the PMT-based likelihood parameters described in Sec.~\ref{sec:PMT_parameter}.

%% file: sec_pure.tex
\section{Reference Scintillation Sample Selection}
\label{sec:sample}
\subsection{$^{22}$Na-calibration data}

It is crucial to procure scintillation signal data devoid of any accompanying noise events, not only for training samples but also for training validation and estimation of selection efficiencies. Consequently, a calibration campaign was conducted over approximately one month utilizing a $^{22}$Na source to collect pure scintillation events.

Most of the $^{22}$Na decays into an excited state of $^{22}$Ne, with 90\% producing positron emission. While the excited $^{22}$Ne emits a 1274\,keV gamma ray to decay to the ground state, the positron annihilates with an electron, resulting in the emission of two back-to-back 511\,keV gamma rays~\cite{BASUNIA201569}. These gamma rays can Compton scatter with electrons in the detector, producing low-energy scintillation signals. The selection criteria for the $^{22}$Na-calibration data require the event coincides with a high-energy event above 200\,keV in another crystal, aligning with the decay scheme.


\begin{figure}[htb]
\centering
\includegraphics[width=.45\textwidth]{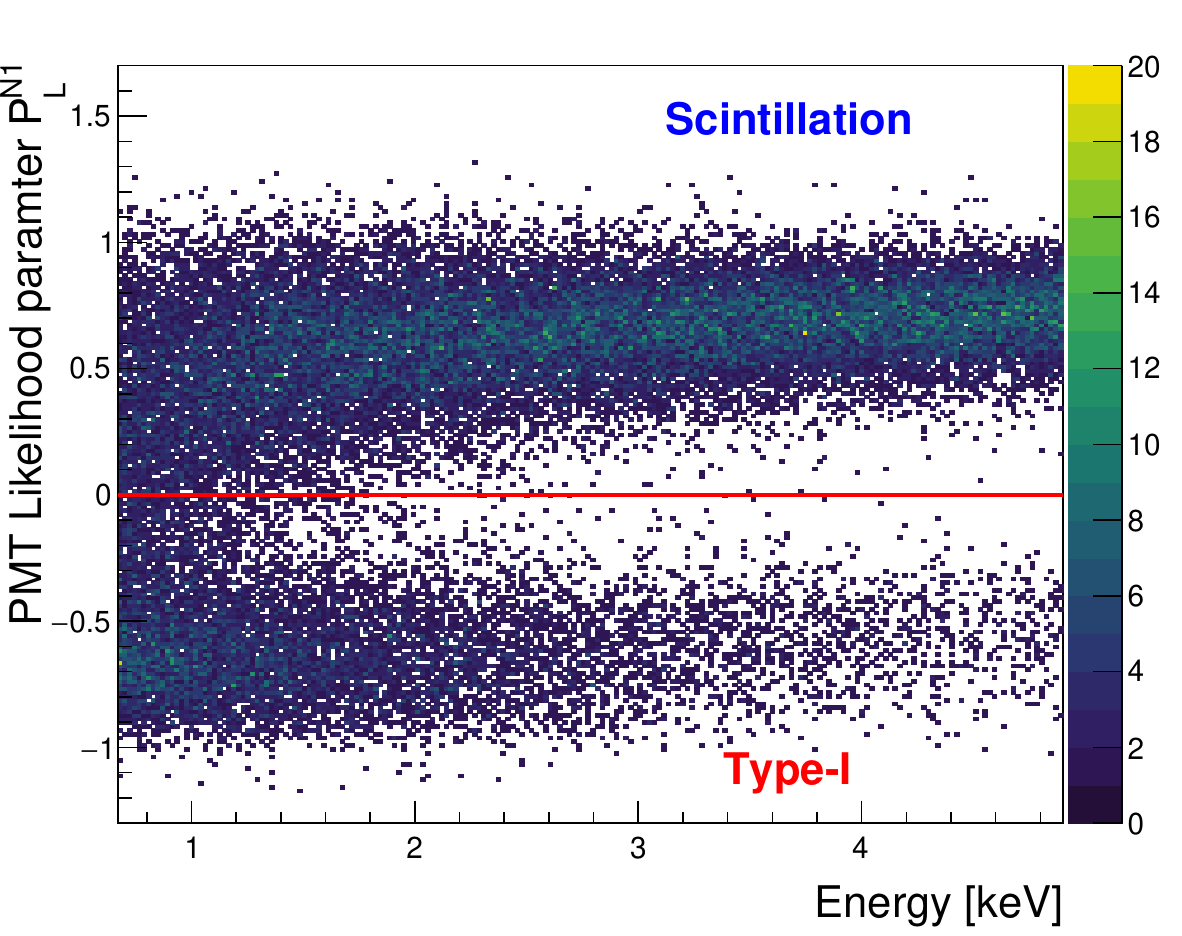}
\qquad
\includegraphics[width=.45\textwidth]{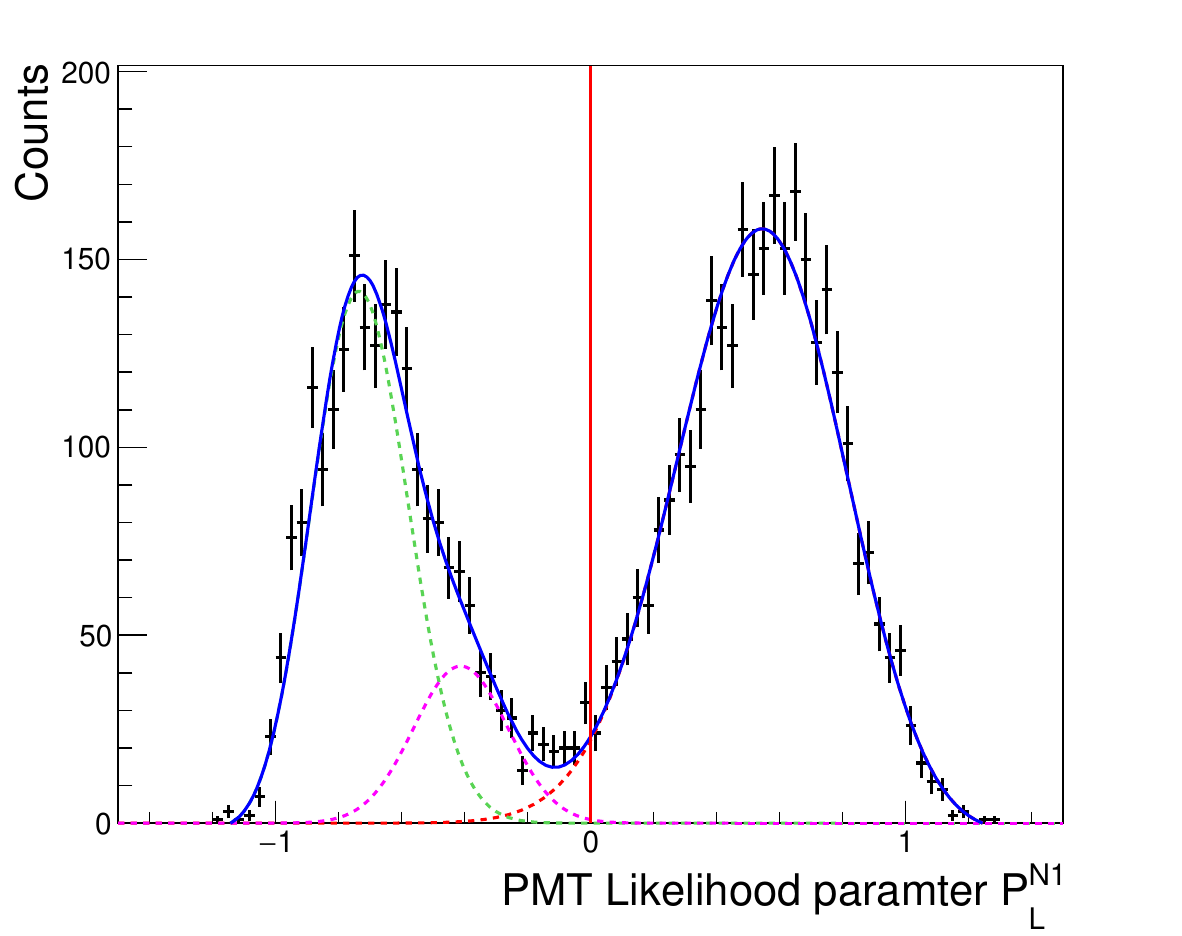}
\caption{(Left) Two-dimensional distribution of PMT-based likelihood parameter for Type-1 noise, $P_\mathcal{L}^{N1}$, versus energy with the calibration data utilizing $^{22}$Na source. (Right) One-dimensional histogram of $P_\mathcal{L}^{N1}$ for the interval [1,1.3]\,keV. The calibration data contain very little Type-2 noise, and the cut depicted by the red line allows the selection of a scintillation-only data set.}
    \label{fig:nasample}  
\end{figure}

Despite these selection criteria, a few PMT-induced noise events, almost entirely Type-1 noise, still remained, what are likely in accidental coincidence with high-energy gamma rays. The distribution of these residual noise events were fairly well characterized using the newly developed parameters in Sec.~\ref{sec:parameter}. Thus we could obtain a sample of the scintillation signal after a cut for $P_\mathcal{L}^{N1}$, as shown in Fig.~\ref{fig:nasample}. 

\subsection{Synchronization of pulse time}
\label{sec:t0correction}
When signals from a crystal cause a trigger, the waveforms of all crystals are recorded, not just the triggered crystal. Thus, waveforms are categorized into two categories: the ``self-triggered'' events are those of the triggered crystal, and ``passively triggered'' events are those recorded because another crystal was triggered. 
While the pulse start times for self-triggered waveforms are aligned closely to 2.4\,$\mu$s, the start times for passively triggered waveforms are often delayed because of a large variation in the arrival times of the first photoelectron for low-energy scintillation events.

Scintillation signal samples from the low-energy events in the $^{22}$Na-calibration data are all passively triggered events due to the selection criteria. Meanwhile, since WIMP-nucleus recoil is such a rare interaction, it is expected to exist only in single-hit events. Consequently, when constructing WIMP-search data, multiple-hit events are vetoed, leaving only single-hit events and thus only self-triggered events.

\begin{figure}[htbp]
\centering
\includegraphics[width=1.0\textwidth]{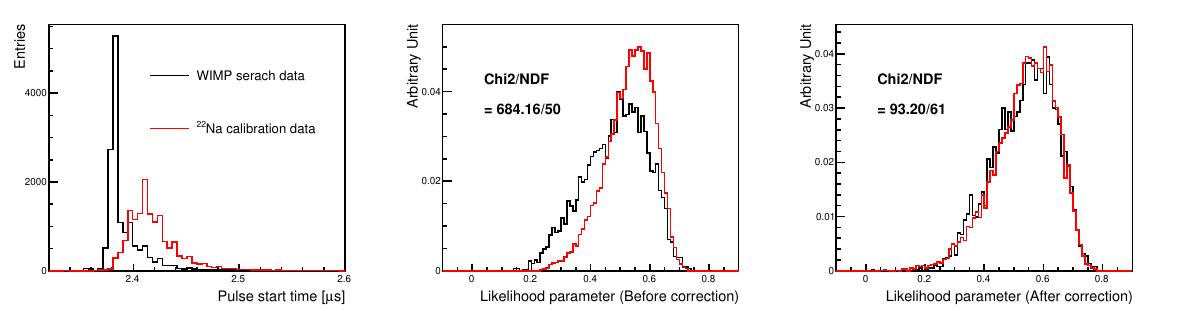}
\caption{(Left)~Comparison of pulse start time distributions between WIMP-search data and $^{22}$Na-calibration data. WIMP-search data mainly consists of self-triggered events, while the calibration data consists of passively triggered events only. Comparison of likelihood parameter distributions (Middle)~before start time correction and (Right)~after start time correction. 
The parameter applying the correction is calculated after aligning the pulse start time of the event to 2.4\,$\mu$s\label{fig:tzcorr}}
\end{figure}

The difference in pulse start time according to trigger type can be found in the left plot of Fig.~\ref{fig:tzcorr}, by comparing distributions between WIMP-search data and $^{22}$Na-calibration data. Since many of the PSD parameters rely on the pulse charge value within a specific time window, this variance in pulse start times can introduce a significant bias in those parameters between self-triggered and passively triggered events. To address this issue, we have aligned the pulse start time of the event to 2.4\,$\mu$s and re-calculated the PSD parameters for event selection. The middle and right plots of Fig.~\ref{fig:tzcorr} displays the comparison of the crystal-based likelihood parameter between $^{22}$Na-calibration data and WIMP-search data, before and after the pulse start time correction. It is clear that after the start time correction, the likelihood parameter of the scintillation calibration data matches much better with that of the WIMP search data. 


%% file: sec_learning.tex
\section{Event Selection via Multi-layer Perceptron Network}
\label{sec:event_selection}

\subsection{Machine learning strategy}
\label{sec:machine_learning}
A multivariate analysis utilizes nine PSD parameters to calculate a score, representing its probability to be a scintillation signal. The newly developed parameters, $P_\mathcal{L}^{N2}$, $F_\mathcal{L}^{N1}$, and $F_\mathcal{L}^{N2}$, have been implemented, while $P_\mathcal{L}^{N1}$ was excluded since it was already utilized in the signal sample selection shown in Fig.~\ref{fig:nasample}.
Additionally, parameters previously developed in Ref.~\cite{COSINE-100:2020wrv} have also been employed. These include the slow and fast charge parameters, the mean-time parameter, and the likelihood parameter $p_l$. 
Two additional parameters with minor modifications to the previously established parameters have also been added. While the original mean-time parameter is calculated with each PMT waveform, the modified mean-time parameter is calculated with the summed waveform.
Furthermore, the original likelihood parameter $p_l$, which was calculated based on 2-ns time binned waveform, has been modified with a calculation based on clustered pulse. The calculation only use the peak-time and the summed-charge of the event's clustered pulses, for the likelihood calculation.


\begin{figure}[htb]
\centering
\includegraphics[width=.45\textwidth]{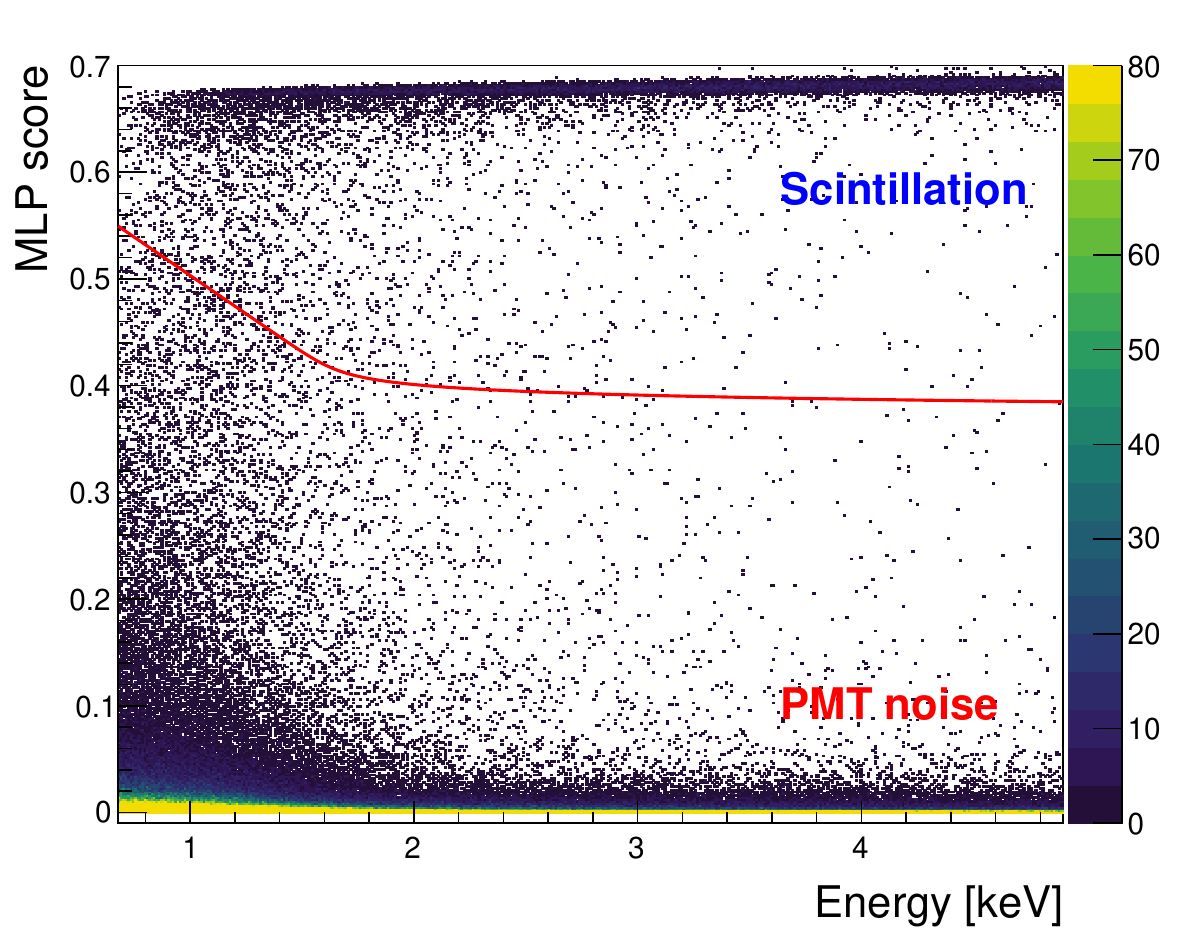}
\qquad
\includegraphics[width=.45\textwidth]{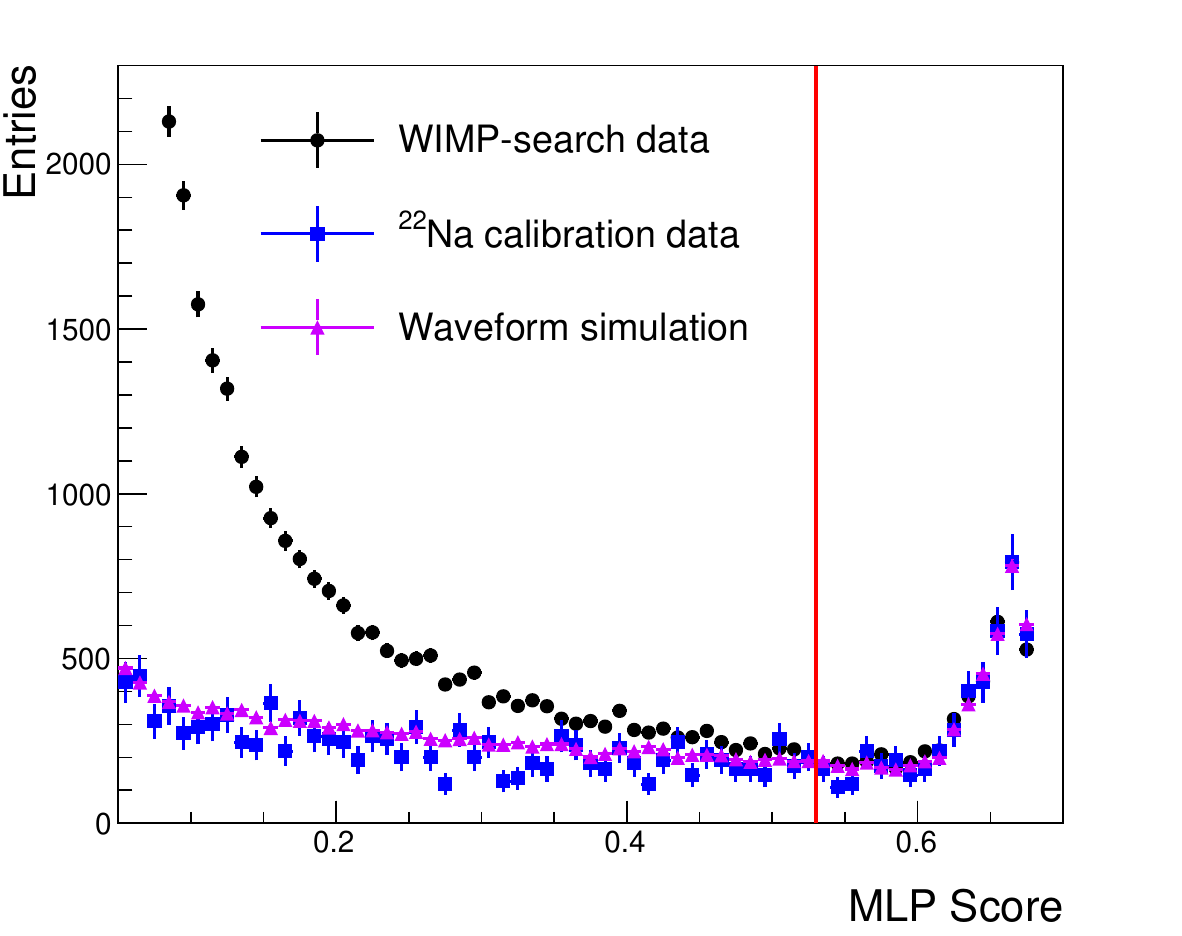}
    \caption{(Left) Distribution of MLP score versus energy, with WIMP-search data. (Right) 1-D histogram of MLP score distribution in the [0.7,1.0]\,keV region. The $^{22}$Na calibration data and waveform simulation are scaled to match entries of WIMP search data in MLP score > 0.63. The red line denotes the cut for a noise contamination rate below 1\%.}
    \label{fig:mlp}  
\end{figure}

For the ML technique, we design a multi-layer perceptron (MLP) having two hidden layers with 32 node each and one output layer. ReLU~\cite{fukushima1975cognitron, nair2010rectified} activation function is adopted for the hidden layers and the Sigmoid~\cite{rumelhart1985learning} activation function is used for the output layer. Also, strategies such as L2 regularization~\cite{ng2004feature} and ADAM optimizer~\cite{kingma2014adam} are used, and the networks are trained by minimizing the cross entropy. After training optimization, we selected a learning rate of $10^{-4}$, batch size of 64, maximum epoch of 2000 and convergence step of 30. The architecture is implemented using the TMVA~(Toolkit for Multivariate Data Analysis with ROOT), an ML training package from CERN ROOT~\cite{root_project,Hocker:1019880,ref_TMVA2}.

During training, the $^{22}$Na calibration data are utilized as the signal sample, while approximately 15\% of the 3-year WIMP search data, containing both scintillation and PMT noise, are used as the background sample. For each crystal, about 800,000 events are used as the training samples. Approximately 0.1\% of the sample consists of signal events, while the majority of the background events are due to PMT noise. We train an MLP network for each crystal separately to account for the difference in waveform characteristics between crystals caused by different crystal histories and PMT behaviors.

As a result, the distribution of the MLP scores as a function of energy is shown in Fig.~\ref{fig:mlp}. The training process does not exhibit over-fitting by comparing the loss functions for the training and test samples.
One can observe in Fig.~\ref{fig:mlp} that the scintillation signal is horizontally distributed at the top and the PMT noise at the bottom, and they are distinguishable down to the region below 1\,keV. A criterion on the MLP score for event selection can be set based on noise contamination using $^{22}$Na-calibration data and WIMP-search data. 
We choose to keep the noise contamination below 1\%, which holds down to the new analysis energy threshold of 0.7\,keV. 

\subsection{Event selection validation}
\label{sec:validation}
Before the newly established 0.7\,keV threshold can be utilized for physics analysis, it needs to be validated in several aspects. First, it must be demonstrated that the events selected from the WIMP-search data are in good agreement with the scintillation samples extracted from the $^{22}$Na-calibration data. In addition, the purity of the $^{22}$Na-calibration data is verified by a comparison with simulations. Finally, since the WIMP-search data was taken over a long period, it must show that the MLP score remains stable over time. 

\begin{figure*}[tb]
    \centering
    \centerline{\includegraphics[width=0.95\linewidth]{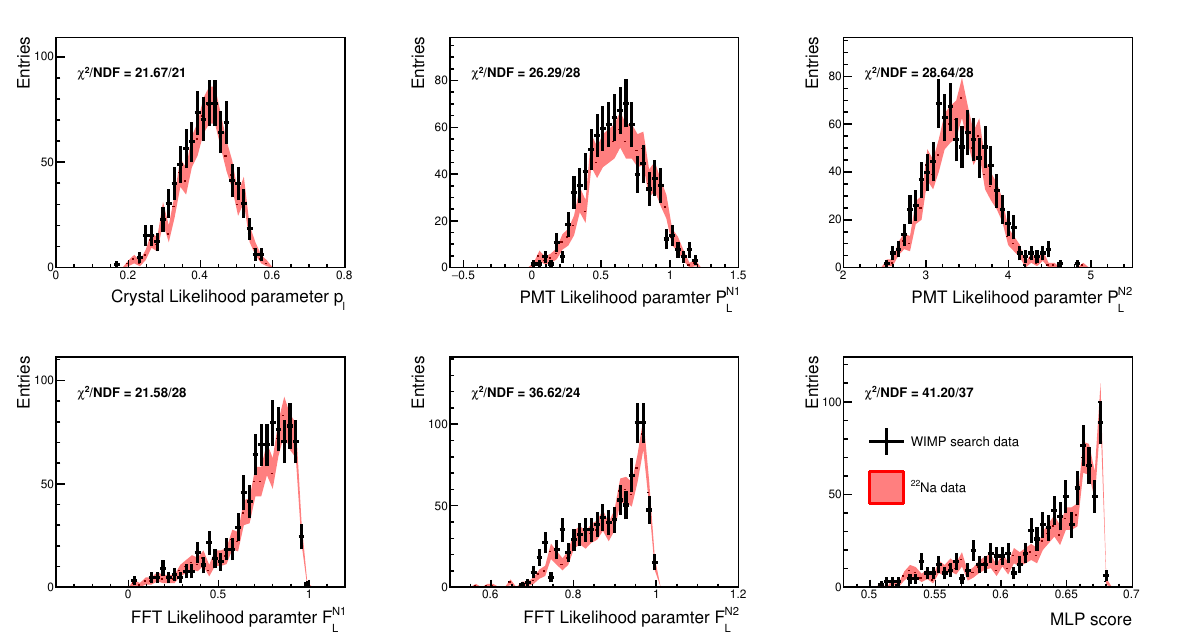}}
    \caption{Comparison between WIMP-search data and 1$\sigma$ band of $^{22}$Na-calibration data in the energy range of [0.7,\,1.0]\,keV. The selection criterion for the MLP score is applied to WIMP-search data only, and $^{22}$Na-calibration data distribution is scaled to WIMP-search data distribution.}
    \label{fig:valid}  
\end{figure*}

To validate the criteria for event selection, WIMP-search data after MLP cut are compared with the $^{22}$Na-calibration data using several PSD parameters. In Fig.~\ref{fig:valid}, the distributions of the $^{22}$Na-calibration data and WIMP-search data for the effective parameters used in training as well as the MLP score are presented, along with the reduced chi-squared values~($\chi^2$/NDF) between two datasets. The reduced chi-squared values are close to 1, indicating that the distributions are consistent with each other.


Despite the utilization of a $^{22}$Na source and the coincidence condition with high-energy events, the purity of the scintillation signal samples used for training and validation may not be 100\%. If the sample is contaminated with even a small amount of noise events, it might affect the criteria for event selection or the estimation of selection efficiencies. To understand the scintillation signal more rigorously, the MLP score distribution of the $^{22}$Na-calibration data is compared with simulation results using a waveform simulation package~\cite{COSINEnPR} developed by the collaboration for analyzing low-energy events.

\begin{figure}[htbp]
\centering
\includegraphics[width=.45\textwidth]{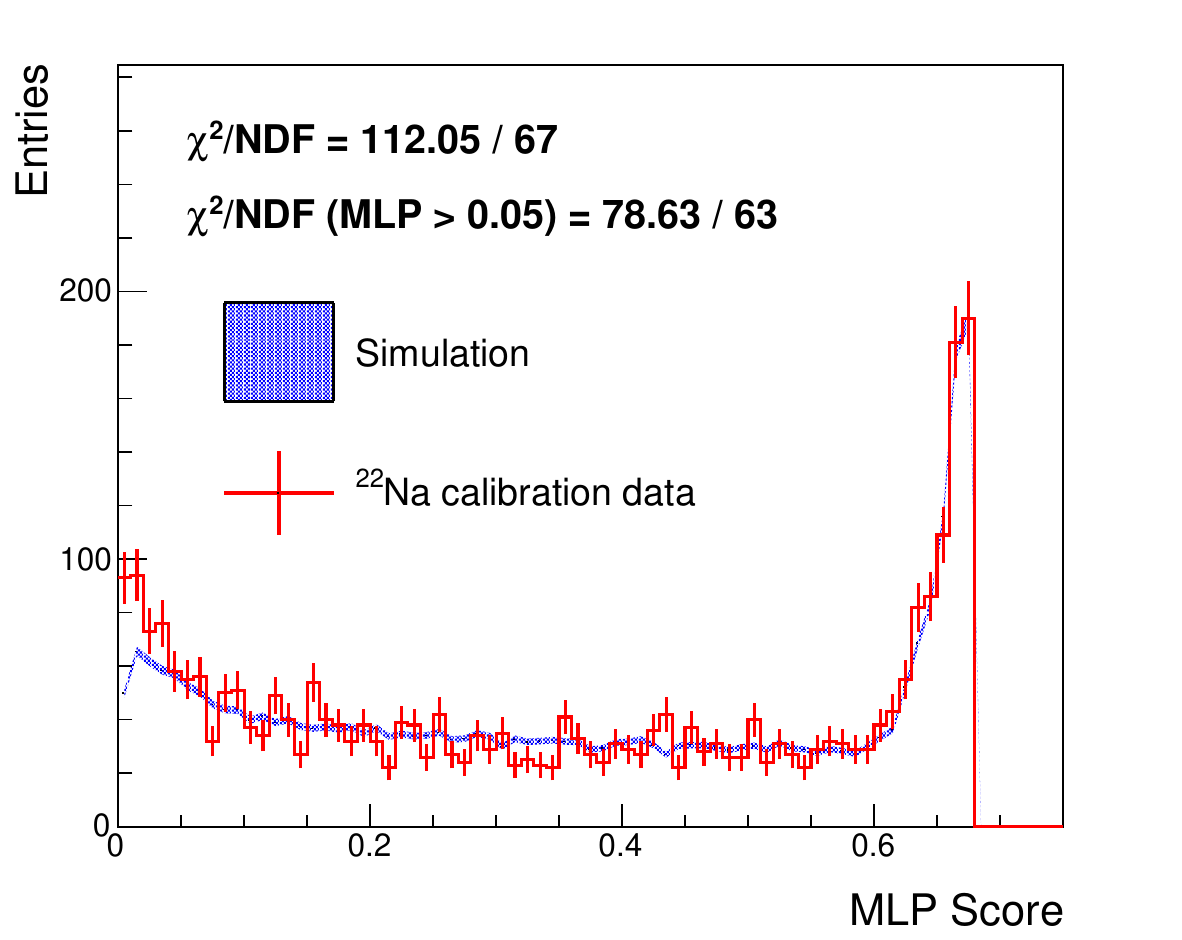}
\qquad
\includegraphics[width=.45\textwidth]{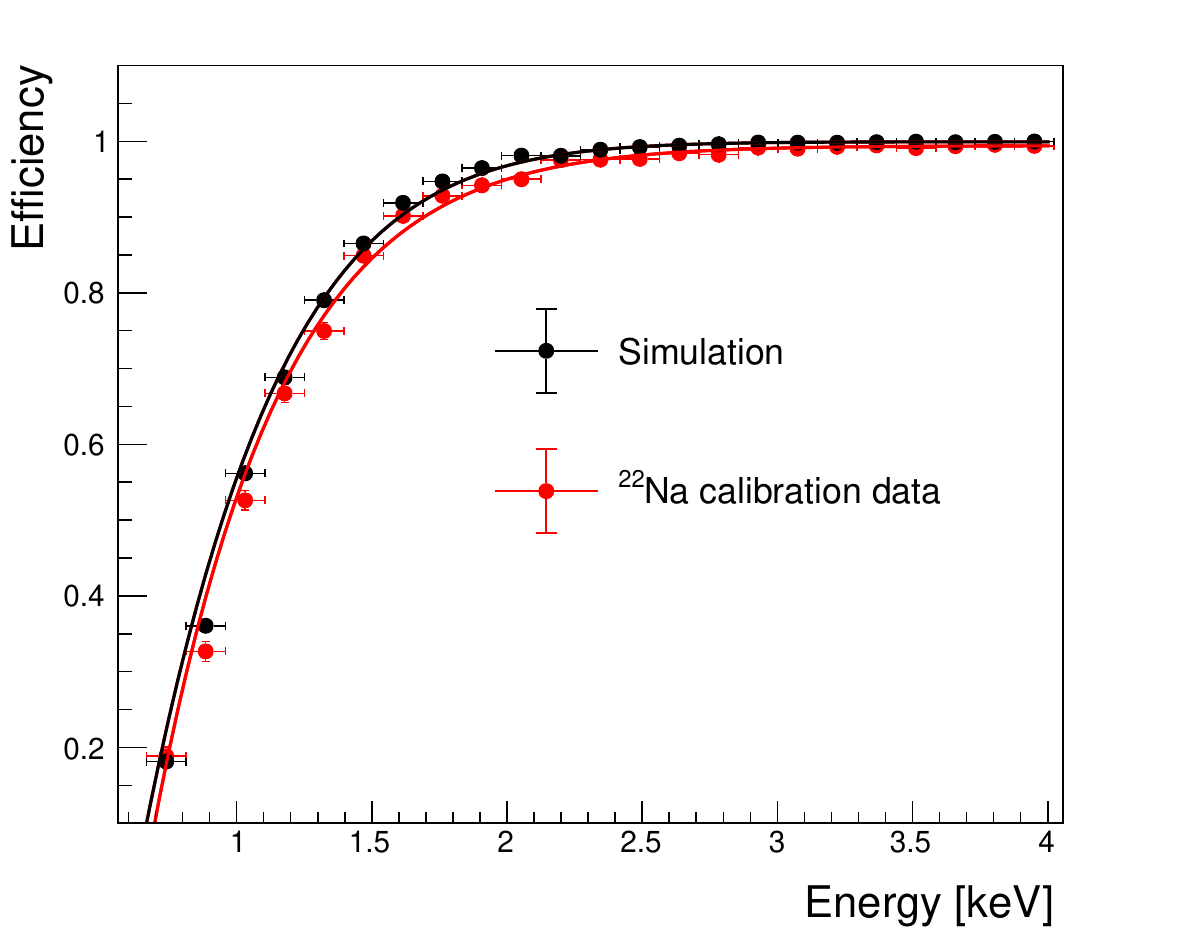}
\caption{(Left) Comparison of MLP score between $^{22}$Na-calibration data and simulation data in the energy range of [0.7,\,1.2]\,keV. Simulation is normalized to $^{22}$Na-calibration data distribution. (Right) Comparison between selection efficiencies estimated from $^{22}$Na-calibration data and simulation data. The energy range is extended to 0.7\,keV from the previous study~\cite{COSINE-100:2020wrv}, and the error bars are propagated from statistical uncertainties.\label{fig:wavesim}}
\end{figure}

The left plot of Fig.~\ref{fig:wavesim} illustrates that the overall MLP distributions of simulation data is within the error bar range of $^{22}$Na-calibration data. One issue is that there is a little bit of excess observed in the low MLP score region. The observed excess is likely due to residual PMT noise events or scintillation events contaminated by the phosphorescence from prior scintillation events. This is not a concern for event selection in the WIMP search energy region, as it can be effectively excluded by the MLP score cut shown in Fig.~\ref{fig:mlp}. When the region with MLP < 0.05 is excluded, the $\chi^2$/NDF value between the $^{22}$Na calibration data and the simulation data approaches 1. 

It is important to note, however, that the excess may give a somewhat inaccurate estimation of the total number of scintillation events. While the number of selected samples is reliable, the inaccuracy of the total number does have an impact on the accuracy of efficiency calculations. The right plot of Fig.~\ref{fig:wavesim} displays the selection efficiencies by MLP cut, indicated by the red line in Fig.~\ref{fig:mlp}, for $^{22}$Na calibration data and simulation. While both exhibit over 10\% efficiencies at around 0.7\,keV, in higher energy region, the simulation-based result shows slightly better efficiency than the values estimated from the $^{22}$Na-calibration data. Consequently, it is concluded that treating the efficiencies estimated via simulation as the mean value and accounting for the difference with the calibration data as a systematic uncertainty would be a prudent approach.

\begin{figure}[htb]
    \centering
    \centerline{\includegraphics[width=0.95\linewidth]{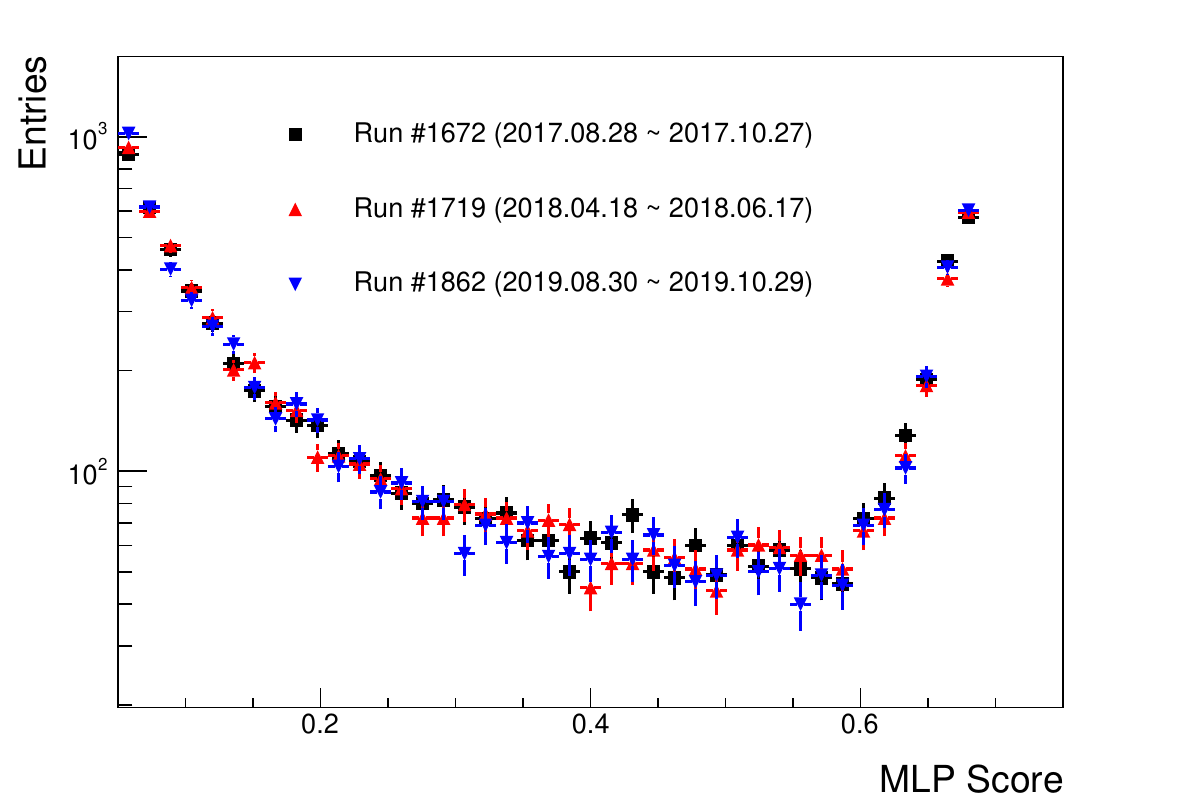}}
    \caption{MLP score distribution for three different time periods. The black, red, and blue dots indicate the early, mid, and late periods of 3-year WIMP-search data, respectively.}
    \label{fig:stability}  
\end{figure}

As a final validation procedure, we examined the MLP distribution in WIMP-search data across three distinct periods. If the MLP distribution varies over time, it could impact the precision of our understanding regarding discrimination power and selection efficiencies. In Fig.~\ref{fig:stability}, we present a comparison of 30-day data for the early, mid, and late data taking periods, demonstrating a consistent MLP distribution within the statistical error range.


%% file: conclusion.tex
\section{Summary} 
\label{sec:summary}
An analysis of event selection in COSINE-100, a direct DM search experiment, is performed, achieving 0.7\,keV analysis energy threshold. To effectively discriminate PMT-induced noise in the low-energy region, new parameters based on the likelihood calculation method have been developed. The newly developed PMT-based and FFT likelihood parameters are demonstrated to be effective and robust discriminators, and the $^{22}$Na source calibration data provide sufficient scintillation signal samples for multivariate MLP training.

As a result of these improvements, the training results of the MLP algorithm allow for an extremely low energy threshold of 0.7\,keV. The MLP training and selection criteria are validated by a comparative analysis of PSD parameters and the selection efficiency of over 10\% at the threshold is confirmed by the waveform simulation. The stability of the MLP distribution over time is also checked, and it was consistent within the statistical uncertainty. This achievement is expected to make a significant contribution to physics analysis, including the search for low-mass WIMPs and the direct test of the DAMA/LIBRA signal.